\documentclass{article}
\usepackage[utf8]{inputenc}
\usepackage{caption}
\usepackage{subcaption}
\usepackage{multicol}
\usepackage{xcolor}
\usepackage[permil]{overpic}
\usepackage{soul}
\usepackage{multirow}

\usepackage{cite}

\newcommand{\sqrtsNN}{\mbox{$\sqrt{\mathrm{s}_{_{\mathrm{NN}}}}$}}
\newcommand{\sqrts}{\mbox{$\sqrt{\mathrm{s}}$}}
\newcommand{\axi}{$\overline{\Xi}^+$}
\newcommand{\xim}{$\Xi^-$}
\newcommand{\alam}{$\overline{\Lambda}$}

\newcommand{\lam}{$\Lambda$}

\newcommand{\omm}{$\Omega^-$}
\newcommand{\aom}{$\overline{\Omega}^+$}

\newcommand{\ppt}{$p_{\rm T}$}

\usepackage{graphicx}
\providecommand{\keywords}[1]{\textbf{\textit{Keywords---}} #1}
\usepackage{amssymb}

\usepackage{graphicx}
\usepackage{multicol}
\begin{document}
%%%%%%%%%%%%%%%%%%%%%%%
%\twocolumn[
%\begin{@twocolumnfalse}

\begin{center}
{\Large \bf Study of Baryon number transport using model simulations in $pp$ collisions at LHC Energies}

\vskip1.0cm
M.~U.~Ashraf$^{1}${\footnote{Email: usman.ashraf@cern.ch (Corresponding Author)}},
Junaid~Tariq$^{2}${\footnote{Email: junaid.tariq@cern.ch}},
Sumaira Ikram$^{3}$,
A.~M.~Khan$^{4}${\footnote{Email: ahsan.mehmood.khan@cern.ch}},
%Khuram Tariq$^{5}$,
Jamila~Butt$^{5}$ and
Samya~Zain$^{6}$\\

{\small\it 
$^1$Pakistan Institute of Nuclear Science and Technology (PINSTECH), Islamabad 45650, Pakistan\\
$^2$Department of Physics, Quaid-i-Azam University, Islamabad 44000, Pakistan\\
$^3$Department of Physics, Riphah International University, Islamabad 44000, Pakistan\\\
$^4$Key Laboratory of Quark \& Lepton Physics (MOE) and Institute of Particle Physics,
Central China Normal University, Wuhan 430079, China\\
%$^5$Shandong University, China\\
$^5$Department of Mathematics and Sciences, Prince Sultan University, Riyadh,  11586, Saudi Arabia\\
$^6$Department of Physics, Susquehanna University, Selinsgrove, PA, 17870, USA}
\end{center}

\vskip1.0cm
%%%%%%%%%%%%%%%%%%%%%%
%\linenumbers
%\date{October 2022}

%\begin{document}

%\maketitle
\begin{abstract}
We report on the excitation function of anti-baryon to baryon ratios ($\overline{p}/p$, {\alam /\lam} and {\axi / \xim}) in $pp$ collisions at {\sqrts} = 0.9, 2.76, 7 TeV from DPMJET-III, Pythia~8, EPOS~1.99, and EPOS-LHC  model simulations. The computed ratios obtained from model simulations are then compared to the experimental results given by the ALICE experiment. To study the predictions of these models at \sqrts = 13.6 TeV, at which LHC is taking Run-3 data in $pp$ collisions, we also computed these ratios at 13.6 TeV. The anti-baryon to baryon ratios are extremely important for the study of baryon number transport mechanisms. These ratios help to determine the carriers of baryon number and in the extraction of baryon structure information. We find that these ratios are independent of both the transverse momentum ({\ppt}) and the rapidity ($y$). Even though all models show a good agreement with experimental data, the ratios extracted from DPMJET-III model closely describes the data at all energies. We found that these ratios converge to unity for various model predictions from 0.9 to 13.6 TeV. It is also observed that, DPMJET-III and EPOS-LHC models exhibits an increase of the ratio with strangeness content for the given energy and this effect is more prominent in {\alam /\lam} and {\axi / \xim} ratios. For all species, the ratio increases with the increase in energy similar to the experimental data. At lower energies, an excess production of baryons over anti-baryons is observed. However, this effect vanishes at higher energies due to the baryon-anti-baryon pair production and the baryon-anti-baryon yield becomes equal. We additionally compute the asymmetry, ($A\equiv \frac{N_{p}-N_{\bar{p}}}{N_{p}+N_{\bar{p}}}$) for protons from various model simulations. The asymmetry shows a decreasing trend with increase in energy from 0.9 to 7 TeV for all energies. This trend is confirmed by model predictions at {\sqrts} = 13.6 TeV, which will help to put possible constraints on model calculations at {\sqrts} = 13.6 TeV once the Run-3 data for LHC becomes available.

\vskip0.5cm
\keywords{simulations, baryon number transport, predictions, pair production, hyperon}
\end{abstract}

%\end{@twocolumnfalse}
%]
%\maketitle
%\begin{multicols}{2}

\section{Introduction}\label{sec1}
Two of the most fundamental physical observables extensively studied and experimentally measured for many decades in both cosmic-ray physics and the hadronic colliders are the yield and the transverse momentum spectra ({\ppt}) of identified particles produced in high energy hadronic interactions~\cite{1}. At collider energies, hadrons at the partonic level are produced both via the hard scattering and the soft scattering processes. Hard scattering is the interaction of two partons with a large momentum transfer  which results in the production of high {\ppt} particles. This process is theoretically explained by the factorization theorem based on perturbative Quantum Chromodynamics (p-QCD) calculations~\cite{2}. Understanding hard scattering processes is important because at the large Hadron collider (LHC), the lower $x$ regime is investigated by increasing the center of mass energy (\sqrts) which results in an increased contribution from hard scattering processes. This increased production of high {\ppt} particles comes from the fragmentation of gluons in the kinematic regions and is probed by the measurements as done in~\cite{2,3}. On the other hand, majority of low {\ppt} ({\ppt} $< 2$ GeV/$c$) particles result from soft scattering processes that have a small amount of momentum transfer. Production of particles in the soft scattering regime cannot be calculated from the first principle and therefore QCD inspired phenomenological models are expected to play a significant role in these calculations. These models are then tuned to be more compatible with prior measurements. Low {\ppt} measurements are  expected to provide additional constraints on simulation models. Furthermore, soft particle production studies are an important tool for understanding basic mechanisms involved in particle production at the low {\ppt} region.

In non-perturbative regimes, calculation of the Quantum Chromodynamics (QCD) processes is not possible with current resources, e.g. it is still unclear in QCD whether the baryon number should be associated with the valence quark or with its gluonic field. However, baryons in QCD are represented by a gauge-invariant state operator which may shed light on what is called the gluonic string junction. A gluonic string junction is a single point in QCD at which three valance quarks join together at a single point~\cite{4, 5}. String junction depicts the association of the baryon number with  baryon gluonic field. It is important to note that in this representation, the string junction is responsible for the production of baryon anti-baryon pairs from the vacuum, whereas the anti-string pair production is responsible for the production of sea quark-antiquarks. The anti-string pair mechanism is responsible for the production of anti-baryons in baryon-baryon collisions.  Another possibility exists that a final baryon may have contributions from string junction or valence quarks, or diquarks of the incoming baryons. The significant diffusion of constituents over the large rapidity results in different baryons and anti baryon spectra at mid-rapidity~\cite{4, 5, 6, 7, 8, 9, 10, 11, 12, 13, 14, 15, 16, 17}.

In Regge field theory, the string junction of baryons coming from the beam at large rapidity $\Delta y$~\cite{18, 19}, can be calculated using the expression, $exp [(\alpha_j - 1) \Delta y]$~\cite{4}, where, $\alpha_j$ is the intercept of the string junction trajectory; $\Delta y  \equiv y_{beam} - y$, and $y_{beam}( \equiv ln(\sqrts / m_B))$ is the rapidity of the incoming baryon; $m_B$ is the mass of the baryon, and $y$ is the rapidity of the string junction. Theoretically, at present, it is not possible to calculate the intercept $\alpha_j$ of the string junction, which is a non-pQCD object.  At mid-rapidity, the (anti-) baryons spectra are expected to vary depending on the value of $\alpha_j$. At $\alpha_j \approx 1$, the string junction of incoming baryons does not show rapidity dependence at high $\Delta y$, as was reported in Ref.\cite{5}. However, at $\alpha_j = 0.5$, transport of string junction approaches zero with the increase in $\Delta y$~\cite{4}.      

Another source that causes differences in the particle and anti-particle spectra is the reggeon exchange with C-parity~\cite{6}, called the asymmetry ($A$), defined as ($A \equiv \frac{N_{p} - N_{\bar{p}}}{N_{p} + N_{\bar{p}}}$). A known Regge pole with intercept $\alpha_{\omega} \approx 0.5$ is called $\omega$-reggeon, whose exchange is one of the main factors which may cause differences in particle and anti-particle spectra. Since the value of $\alpha_{\omega}$ is less than one, its contribution decreases at mid-rapidity with the increasing colliding energy. Another source that causes differences in the particle and anti-particle spectra at mid-rapidity region, if it exists, is the negative signature of Regge pole and $\alpha_j \approx 1$. 

The information about contributions of different mechanisms involved in the study of baryon production from baryon and anti-baryon spectra in $pp$ collisions can be gathered. Particularly, the measurement of the anti-baryon to baryon spectra and $\overline{B}/B$ ratios with different quark flavors, for example $p (\overline{p})$, {\lam (\alam)}, {\xim(\axi)}, and {\omm (\aom)} are directly used to find constraints on the mechanisms involved in the baryon production. %\textcolor{blue}{THIS IS OUR RESULT - SHOULD WE HAVE IT IN HERE????--- For example, increase in the strangeness content results in the reduced contribution of the stopping related processes of different beam constituents. As a result, the $\overline{B}/B$ ratio becomes closer to unity as the strangeness content increases.} 

In this paper, we present the excitation function of the baryons at {\sqrts} = 0.9, 2.76, 7 TeV which is represented by the energy dependence of baryon ratios. To perform this task a comprehensive study was undertaken utilizing events generated from the various model-based event generators, namely the DPMJET-III, Pythia~8, EPOS~1.99, and EPOS-LHC. These events were used to study different $\overline{B}/B$ ratios ($\overline{p}/p$, {\alam /\lam}, {\axi / \xim}) at {\sqrts} = 0.9, 2.76, 7 TeV. The results thus obtained are then compared to the published data from the ALICE experiment~\cite{20}. We also performed simulations at {\sqrts} = 13.6 TeV to check the model predictions, where no current experimental data is available. However, experimental data should be available later because LHC is currently taking data at {\sqrts} = 13.6 TeV in Run-3. 

This paper is organized as follows: a description of models in section~\ref{sec2}, followed by analysis, results and discussion presented in section~\ref{sec3}. Finally, the summary and conclusion presented in section~\ref{sec4}.

\section{Model Details}\label{sec2}
In this section, the models used for simulation are discussed in detail. 
\begin{itemize}
    \item \textbf{DPMJET-III}: DPMJET-III unifies all features of the DPMJET-II~\cite{22}, Dtunuc-2~\cite{23, 24} and PHOJET-1.12~\cite{25} event generators into a single coding system. DPMJET model is based on multiple scattering Gribov-Glauber formalism and the  simulations of hadron-nucleus ($hN$), hadron-hadron ($hh$), photon-hadron ($\gamma h$), nucleus-nucleus ($NN$), photon-nucleus ($\gamma N$) and photon-photon ($\gamma \gamma$)  interactions can be studied from few GeV to the highest cosmic ray energies~\cite{26}. 
    DPMJET is based on the Dual Parton Model (DPM) and it describes the soft and multi-partonic interactions in high-energy interactions~\cite{26}. The soft processes are defined by the exchange of pomerons under the Reggeon field theory~\cite{27}, whereas, hard processes are described by the perturbative parton scattering approach. The assumptions of duality~\cite{28} with Gribov's Reggeon field theory~\cite{27} and the predictions of large number of flavors ($N_f$) and number of colors ($N_c$) expansions of QCD~\cite{29} are combined in the DPM. The production cross-sections, as well as the total (quasi) elastic calculations for various colliding systems, can be calculated under the framework of DPMJET at high energies~\cite{30}. The new feature that uses enhanced graph cuts in the non-diffractive inelastic $hN$ and $NN$ collisions is incorporated in DPMJET-III model. On the other hand, the hadronization of color-neutral strings is based on the Lund model as incorporated in Pythia~\cite{31, 32}. Further details of the DPMJET-III model can be found in Ref.~\cite{26}.

\item \textbf{Pythia 8}: Pythia 8~\cite{26, 31, 32, 33} simulates  in detail the particle production in high energy collisions over a wide range of energy scales. Due to the complexity of the hadron collisions and production, no one comprehensive theory can predict the event properties over the full range of available collision energies. Pythia 8 with a core, based on the Lund string model of hadronization~\cite{34}, addresses a large set of phenomenological problems in particle, astroparticle, nuclear, and neutrino physics. According to the string model,  QCD-vacuum expels a color flux when two color charges are located at a large distance within it. This leads the development of a color-flux tube between color charges which behaves as a string. In string fragmentation, these strings are then broken to form hadrons. Pythia 8 also includes a hybrid hadronization model~\cite{35} to accommodate heavy-ion collisions. For the historical evolution of the Pythia event generator, see Ref.~\cite{36}.

Pythia 8 can simulate many Standard-Model processes, that include lepton-lepton, lepton-hadron and hadron-hadron, Beyond Standard Model~(BSM) particle decays, resonance decays with hadronization and final state showering, hadronization of partonic configuration, $NN$ collision for $\sqrtsNN>10~$GeV and astrophysical phenomenons. Pythia includes a large selection of QCD processes classified into three groups: (1) interaction of light quarks and gluons,~2~$\rightarrow$~2~ (i.e, gg~$\rightarrow$~gg,~gg~$\rightarrow~q\bar{q}$~), (2) production of charm and bottom,~2~$\rightarrow$~2~(~i.e.,~gg~$\rightarrow~c\bar{c}$,~gg~$\rightarrow~b\bar{b}$~) and (3) processes involving light quarks and gluons, ~2~$\rightarrow$~3. Electroweak processes in Pythia include prompt photon production~($q$g~$\rightarrow~q\gamma$,~gg~$\rightarrow$~g$\gamma$~), weak boson, single vector boson~($\gamma^{*}/Z$,~$W^{\pm}$), photon collision~(~$\gamma \gamma\rightarrow q\bar{q},~c\bar{c},~b\bar{b}$~). Pythia 8 also provides charmonium and bottomonium production using non-relativistic QCD~(NRQCD)~\cite{37} including both color singlet and color octet configuration, top, Higgs production, and supersymmetric particles using Minimal Supersymmetric Simplified Model~(MSSM)~\cite{38}.

\item \textbf{EPOS-1.99/LHC}~\cite{39,40,41}: 
In simple hadron-hadron interaction models at high energies, the inclusive cross-section is the convolution of the two
parton distribution function~(PDF)~\cite{42}. Where the inclusive cross-section is deduced from the pQCD and PDF from deep inelastic scattering experiments. In the EPOS model, emission of partons after hard scattering is called the initial cascade or space-like cascade. These partons are generally off-shell and give rise to parton emission called final cascade or time-like cascade. The cascade of partons, including initial and final cascade, is called a parton ladder. Parton ladders are split into color strings which then fragment into hadrons. The soft part of the parton ladder is parameterized in Regge pole fashion~\cite{43}. To complete the picture, remnants which are usually colorless excited quark-antiquark are also taken into account. Thus in EPOS model, the hadron-hadron interaction is composed of two parts, inner contribution (from the parton ladder) and outer contribution (from the remnants). Remnants produce particles at large rapidities and parton ladders at central rapidities.\\
EPOS~1.99~\cite{41}, which was released in 2009 was adjusted specifically to allow for an in-depth study of LHC data. This version was called EPOS-LHC~\cite{40}. EPOS-LHC can be adjusted (tuned) to reproduce different hadronic interactions, that include, $pp$, $pN$, and $NN$ interactions, where $N$ can be 1 to 210 nucleons in the energy range from 40 GeV in the lab frame to 1000 TeV in the center-of-mass frame. Collective hadronization in $pp$ collision produced by the very high density matter at the LHC energies are resolved by the introduction of various sets of parameters about flow effects in EPOS-LHC. Such highly dense matter core expands very quickly, requiring a unique radial flow algorithm used in EPOS-LHC.  
\end{itemize}

\section{Analysis, Results and Discussion}\label{sec3}
In this section, we  discuss the data analysis and results in detail. We start the discussion with the introduction to the data sets utilized for this paper in section~\ref{sec3-1}. 

\subsection{Data Set}\label{sec3-1}

For data analysis, 35 million events for $pp$ collisions were simulated for each model at {\sqrts} = 0.9, 2.76 and 7 TeV using the DPMJET-III, Pythia~8, EPOS~1.99, and EPOS-LHC models. LHC is currently running (Run-3) to take high luminosity data from $pp$ collisions at {\sqrts} = 13.6 TeV. In order to study the predictions of different models, we also simulated 30 million events for $pp$ collisions at {\sqrts} = 13.6 TeV. The simulations are then compared to the available published results from the ALICE experiment~\cite{20}. Motivated by the inherent limitations of the current ALICE experiment, the final state hyperons are considered in the rapidity window $|y| < 0.8$ and $|y| < 0.5$ in case of $\overline{p}/p$ ratios. The particles with lifetime $c\tau >$ 10 mm are considered to be final state particles because they can be tracked in the detector. The cuts used in this study to calculate the transverse momentum ({\ppt}) and rapidity ($y$) distributions are listed in Table 1.   

%%%%%%%%%%%%%%%%%%%%%%%%%%%%%%%%%%%%%%%%%%%%%%%%%%%%%%%%%%%%%%%%%%%%%
%   This is a LaTeX file.
%%%%%%%%%%%%%%%%%%%%%%%%%%%%%%%%%%%%%%%%%%%%%%%%%%%%%%%%%%%%%%%%%%%%%

%\documentclass[10pt]{article}
%\usepackage{multicol}
%\usepackage{graphicx}
%\usepackage{amsmath}
%\usepackage[a4paper]{geometry}
%\usepackage{rotating}

%\renewcommand{\baselinestretch}{1.05}
%\renewcommand{\thefootnote}{\fnsymbol{footnote}}
%\setlength{\parindent}{.5cm} \setlength{\columnsep}{.5cm}
%\setlength{\oddsidemargin}{-.5cm} \setlength{\topmargin}{-1.5cm}
%\setlength{\textwidth}{17.5cm} \setlength{\textheight}{23.5cm}

%\begin{document}

\begin{table*}[!htb]
{
\begin{center}
\caption{{\ppt} and $y$ cuts used to study $\overline{B}/B$ ratios at LHC energies}
\begin{tabular}{|c||c|c||c|c||c|c|}
\hline\hline
  %  \multicolumn{2}{c}{Multi-column}
   $\sqrt{s}$ (TeV)&  \multicolumn{2}{c||}{$\overline{p}/p$} & \multicolumn{2}{c||}{{\alam /\lam}} & \multicolumn{2}{c|}{{\axi / \xim}} \\
\hline\hline
 & $p_T$ & $|y| $ &  $p_T$ & $|y| $ & $p_T$ & $|y| $ \\
\hline
     0.9   &  ]0.45 , 1.05[  &$  < 0.5$   & ]0.5 , 4.0[ &$ < 0.8$    & ]0.5 , 3.5[ &   $< 0.8$  \\
\hline
     2.76   &  ]0.45 , 1.05[  &$  < 0.5$   & ]0.5 , 4.5[ &$ < 0.8$    & ]0.5 , 4.5[ &   $< 0.8$  \\
\hline
     7   & ]0.45 , 1.05[  &$  < 0.5$   & ]0.5 , 10.5[ &$ < 0.8$    & ]0.5 , 5.5[ &   $< 0.8$  \\
 \hline
     13.6   & ]0.45 , 1.05[  &$  < 0.5$   & ]0.5 , 10.5[ &$ < 0.8$    & ]0.5 , 5.5[ &   $< 0.8$  \\

   %  & $0.45 < p_T < 1.05$ GeV/$c$ & $0.5 < p_T < 4.0$ GeV/$c$ & $0.5 < p_T < 3.5$ GeV/$c$ \\
     %2.76   & $|y| < 0.5$   & $|y| < 0.8$    &  $|y| < 0.8$  \\
     %& $0.45 < p_T < 1.05$ GeV/$c$ & $0.5 < p_T < 4.5$ GeV/$c$ & $0.5 < p_T < 4.5$ GeV/$c$ \\
     %7   & $|y| < 0.5$   & $|y| < 0.8$    &  $|y| < 0.8$  \\
    % & $0.45 < p_T < 1.05$ GeV/$c$ & $0.5 < p_T < 10.5$ GeV/$c$ & $0.5 < p_T < 5.5$ GeV/$c$ \\
     %13.6   & $|y| < 0.5$   & $|y| < 0.8$    &  $|y| < 0.8$  \\
     %& $0.45 < p_T < 1.05$ GeV/$c$ & $0.5 < p_T < 10.5$ GeV/$c$ & $0.5 < p_T < 5.5$ GeV/$c$ \\
    
\hline
\end{tabular}
\label{tab1}
\end{center}}
\end{table*}

%\end{document}

\subsection{Results}

We report the $\overline{B}/B$ ($\overline{p}/p$, {\alam /\lam} and {\axi / \xim}) ratios calculated utilizing the various model simulations discussed in section~\ref{sec2} as a function of {\ppt} and $y$ at the LHC energies. It is important to note that due to insufficient statistics, the {\aom / \omm} ratio was not studied in this analysis. The results of each of the $\overline{B}/B$ ratios are reported in detail below.

\begin{figure}[!ht]
\centering
\includegraphics[width=0.49\textwidth,height=0.35\textheight]{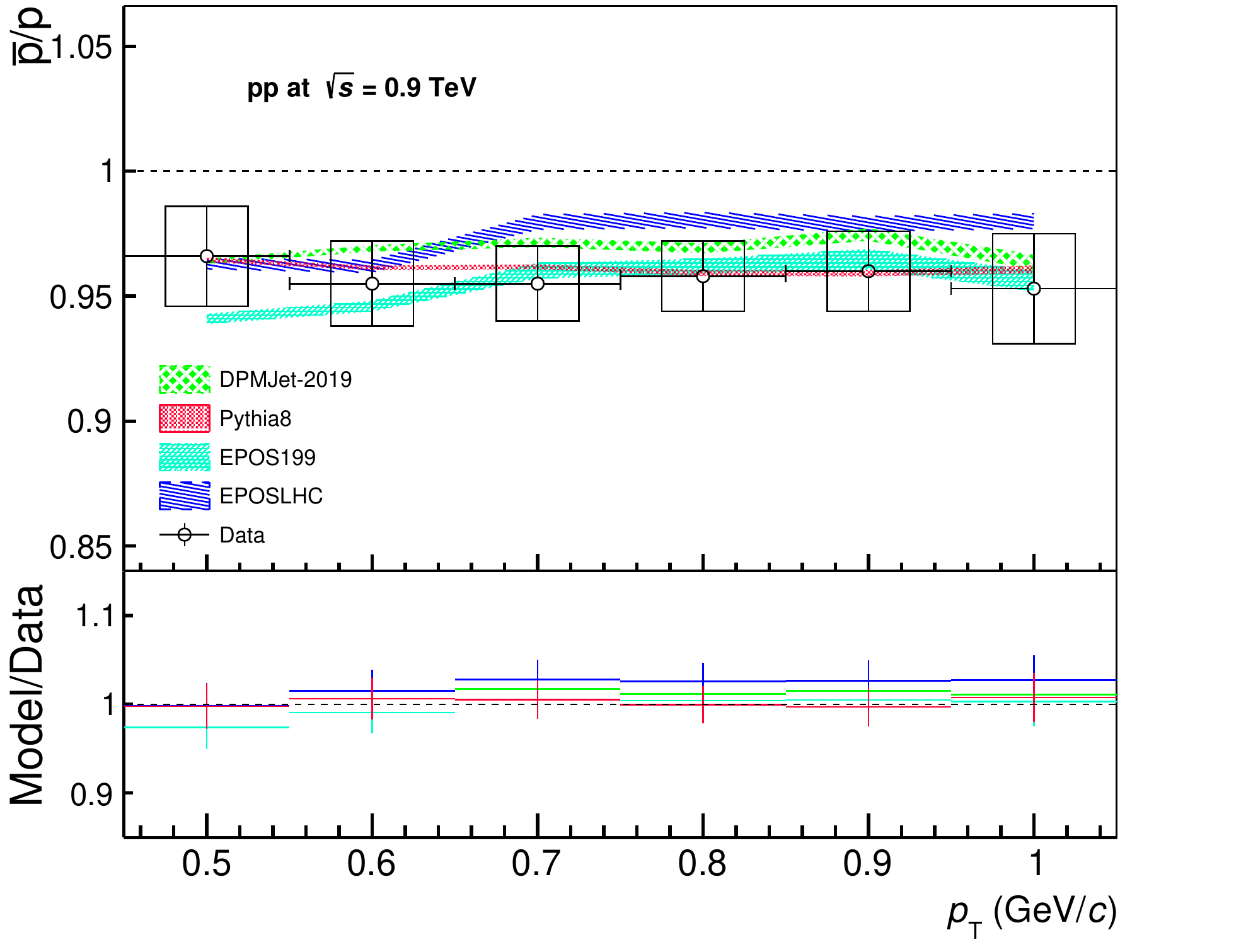}

\caption{{\ppt} dependent $\overline{p}/p$ ratio in $pp$ collisions at {\sqrts} = 0.9 TeV compared with DPMJET-III, Pythia~8, EPOS~1.99, and EPOS-LHC models. The model-to-data ratio shown in the lower panel. }
\label{fig1}
\end{figure}

 The $\overline{p}/p$ ratio vs {\ppt} in $pp$ collisions at {\sqrts} = 0.9 TeV is shown in fig.~\ref{fig1}. The simulation results calculated from DPMJET-III, Pythia~8, EPOS~1.99, and EPOS-LHC models are then compared with the experimental data from ALICE~\cite{20}. The data from ALICE shows no {\ppt} dependence and $\overline{p}/p$ ratio is almost smooth around 0.95 across all {\ppt} bins. In comparison, the DPMJET-III, Pythia~8, and EPOS~1.99 models successfully reproduce the data trends and show no {\ppt} dependence. On the other hand, the EPOS-LHC model slightly over predicts the ratio at higher {\ppt} bins ( $p_T > 0.7$ GeV/$c$). This agreement is further confirmed by the model-to-data ratio as shown in the lower panel of fig.~\ref{fig1}.

\begin{figure}[!ht]
\centering
\includegraphics[width=0.49\textwidth,height=0.35\textheight]{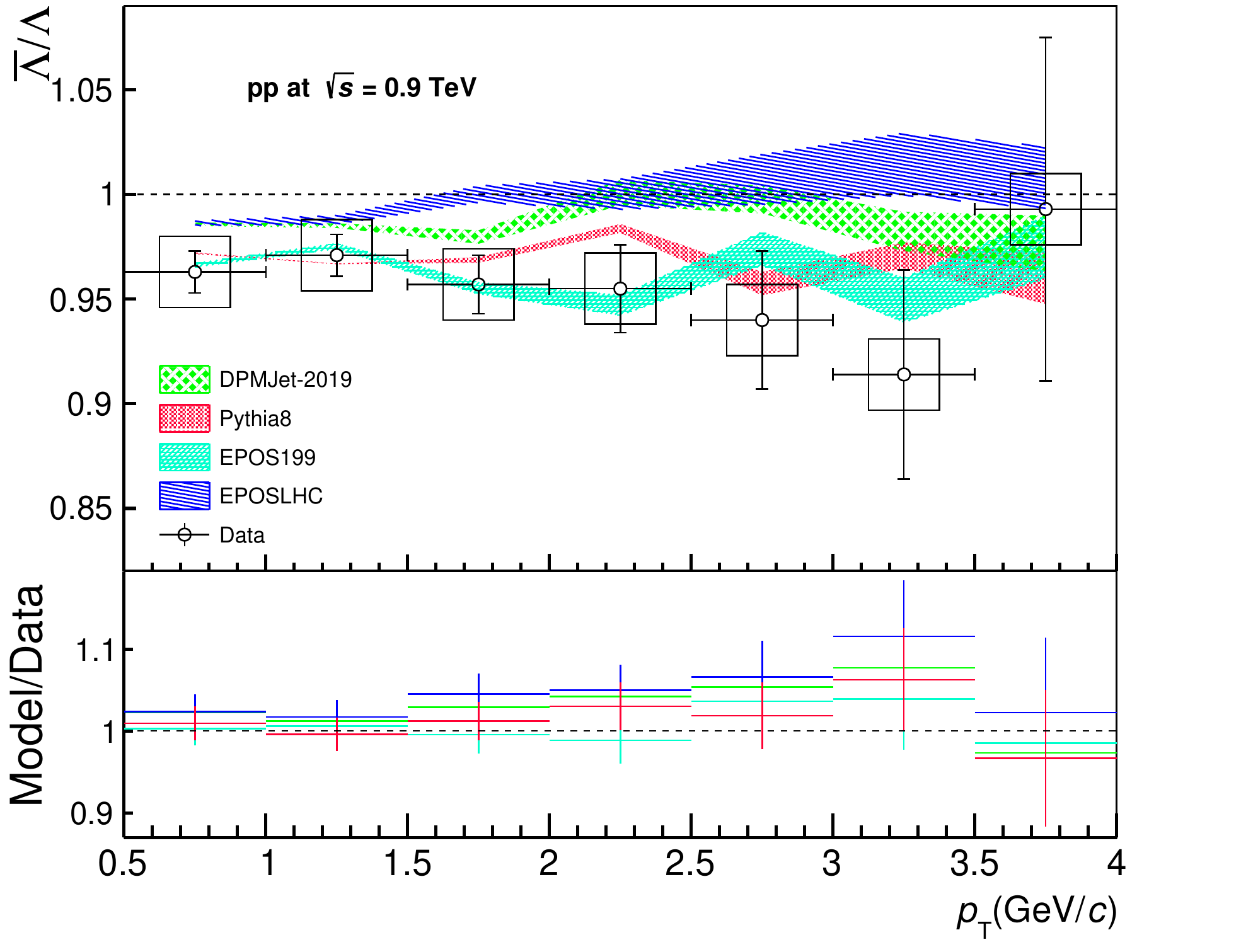}
\includegraphics[width=0.49\textwidth,height=0.35\textheight]{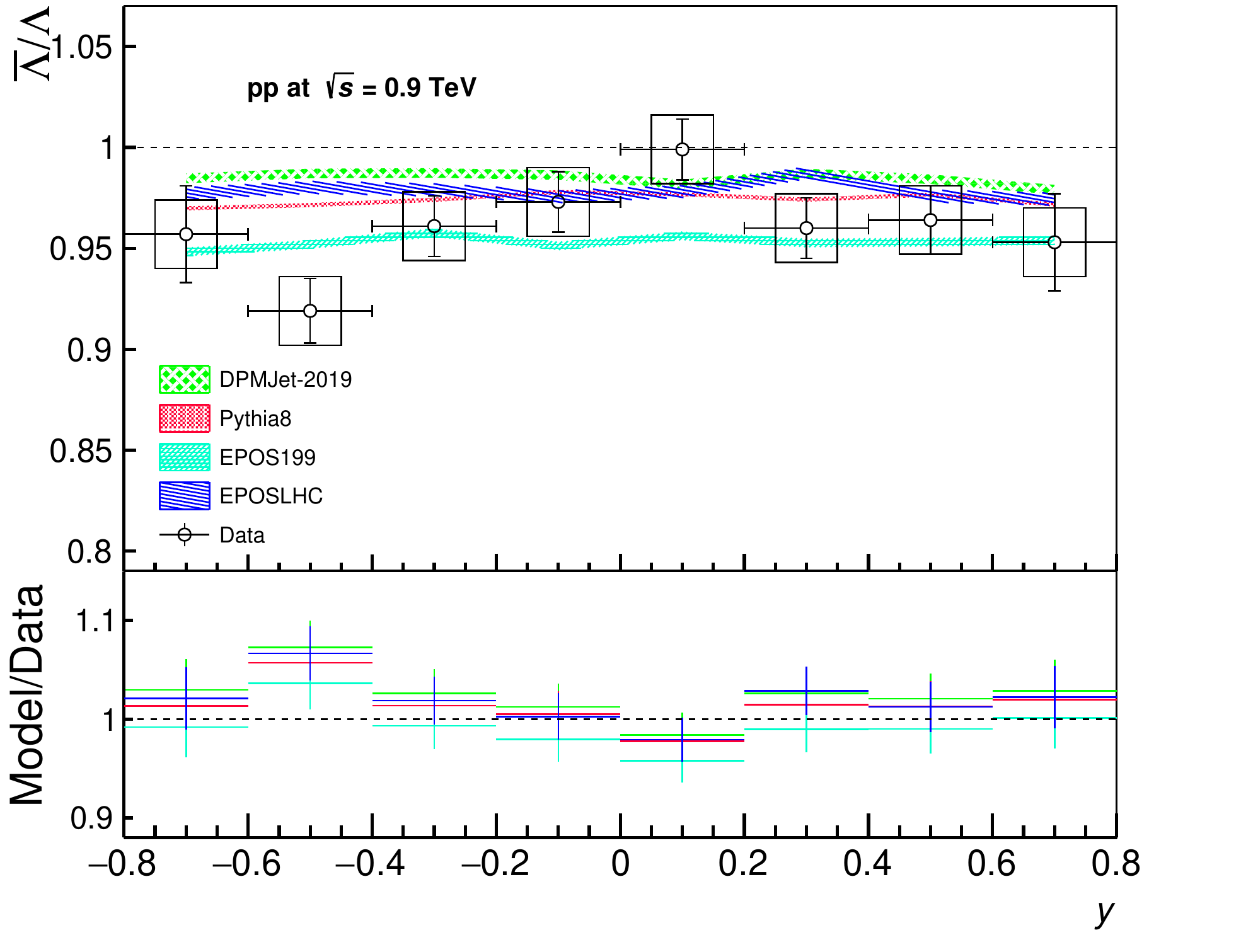}

\caption{{\ppt} dependent {\alam /\lam} ratio (left) and $y$ (right) in $pp$ collisions at {\sqrts} = 0.9 TeV compared with DPMJET-III, Pythia~8, EPOS~1.99, and EPOS-LHC models. The model-to-data ratio shown in the lower panel. }
\label{fig2}
\end{figure}

Figure~\ref{fig2} (left) represents the ({\alam /\lam}) ratio vs {\ppt} in $pp$ collisions at {\sqrts} = 0.9 TeV from the ALICE experiment and the results calculated from the model simulations. The ALICE data  shows a slight {\ppt} dependence for {\ppt}$>$ 2 GeV/$c$. It is observed that EPOS~1.99 and Pythia~8 are in good agreement with experimental data from ALICE within uncertainties. The DPMJET-III model at low {\ppt} bins ($<$ 2 GeV/$c$), reasonably reproduces the data while it over predicts the data at intermediate {\ppt} bins. The EPOS-LHC model clearly over predicts data for all {\ppt} bins except for {\ppt} ($>$ 3.5 GeV/$c$)) at which the model matches the data within uncertainties. The lower panel of fig.~\ref{fig2} (left), gives model-to-data ratio which shows good agreement of model simulations with experimental data. For EPOS-LHC the model-to-data ratio shows $<$ 10\% difference apart from the $[3-3.5]$ GeV/$c$ {\ppt} bin for which the difference is about 12\%. 

Figure~\ref{fig2} (right) shows the ({\alam /\lam}) ratio as a function of $y$ in $pp$ collisions at {\sqrts} = 0.9 TeV from the ALICE experiment compared to the model simulations. It is observed that all models successfully reproduce the ({\alam /\lam}) ratio as a function of $y$ for most of the $y$ bins with model-to-data difference of $<$ 10\%. Overall, we find that all models are independent of the $y$ and closely describe the ({\alam /\lam}) ratio for all $y$ bins.

\begin{figure}[!ht]
\centering
\includegraphics[width=0.49\textwidth,height=0.35\textheight]{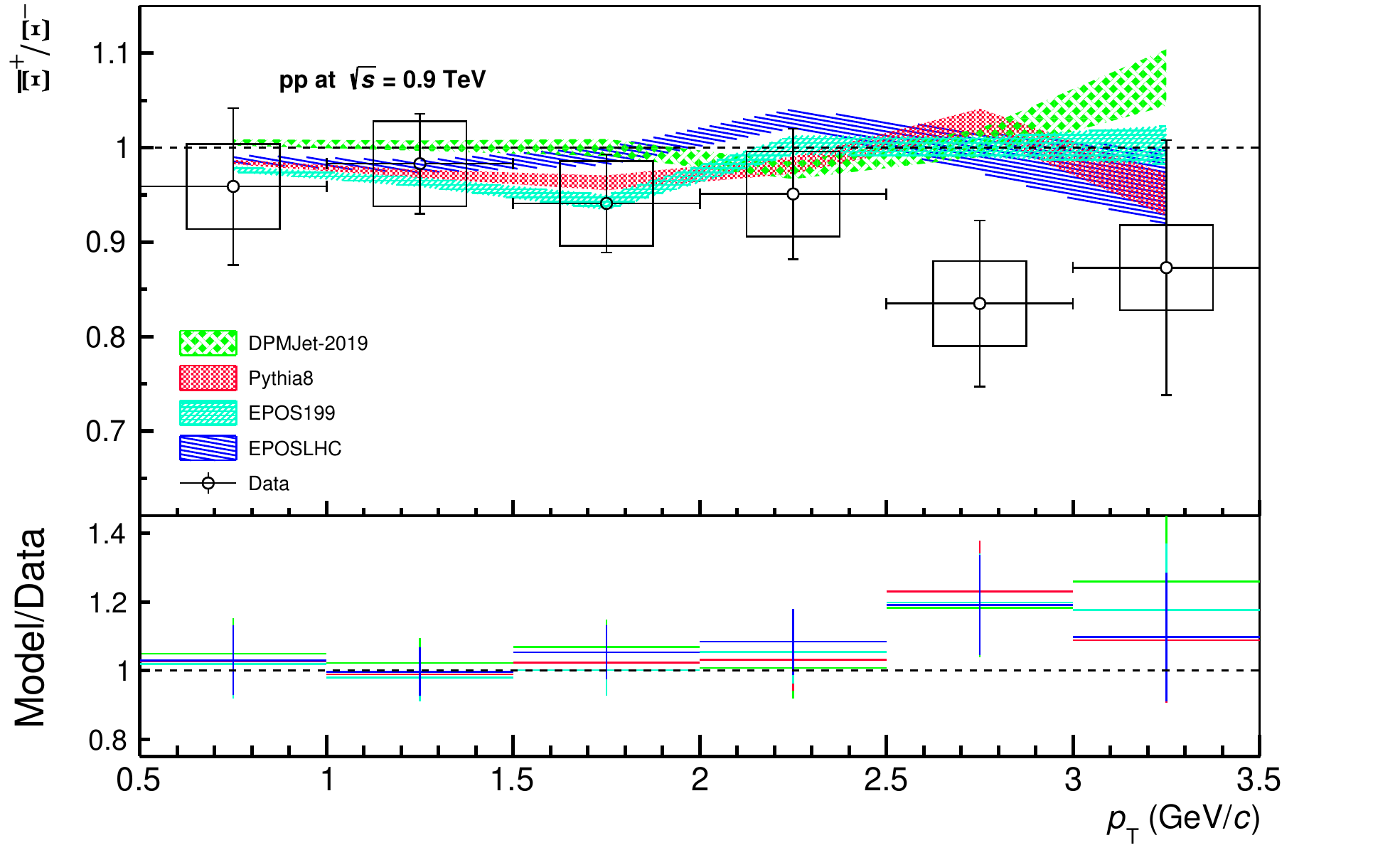}
\caption{{\ppt} dependent {\axi / \xim} ratio in $pp$ collisions at {\sqrts} = 0.9 TeV compared with DPMJET-III, Pythia~8, EPOS~1.99, and EPOS-LHC models. The model-to-data ratio shown in the lower panel. }
\label{fig3}
\end{figure}

Figure~\ref{fig3} shows the comparative data for $pp$ collisions at {\sqrts} = 0.9 TeV from ALICE experiment and the model simulations for the ({\axi / \xim}) ratio as a function of {\ppt}. For ALICE data the ({\axi / \xim}) ratio is almost unity for all {\ppt} bins except for {\ppt} $>$ 2.5 GeV/$c$ where there is a sudden decrease in the ratio. The model simulations match experimental data within uncertainties for all {\ppt} bins except $[2.5 < p_\mathrm{T} < 3.0]$ GeV/$c$ for Pythia~8, EPOS~1.99 and EPOS-LHC, and {\ppt} $>2.5$ GeV/$c$ for DPMJET-III . The ({\axi / \xim}) ratio computed from nearly all models is around unity for all {\ppt} bins. Additionally, the observed model-to-data ratio is shown in the lower panel of fig.~\ref{fig3}, where the uncertainty in the last two bins ({\ppt} $>$ 2.5 GeV/$c$) is around 20\%. Apart from these two bins, the model simulations are in good agreement with the experimental data within statistical uncertainties.

\begin{figure}[!ht]
\centering
\includegraphics[width=0.49\textwidth,height=0.35\textheight]{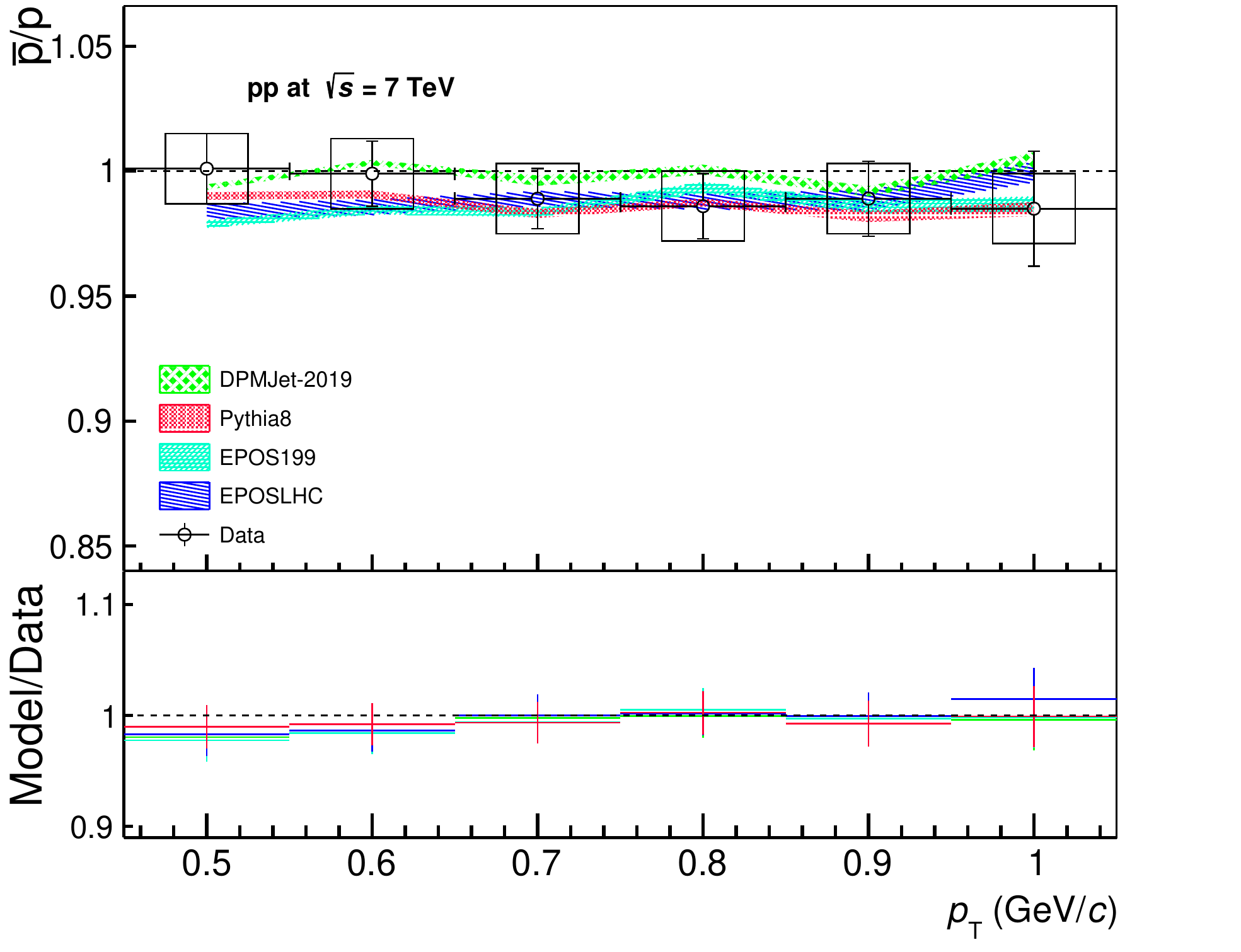}
\caption{{\ppt} dependent $\overline{p}/p$ ratio in $pp$ collisions at {\sqrts} = 7 TeV compared with  DPMJET-III, Pythia~8, EPOS~1.99, and EPOS-LHC models. The model-to-data ratio is shown in the lower panel. }
\label{fig4}
\end{figure}

The ($\overline{p}/p$) ratio vs {\ppt} for ALICE data in $pp$ collisions at {\sqrts} = 7 TeV along with model simulations are given in fig.~\ref{fig4}. Neither the experimental nor the simulated data show any {\ppt} dependence and the ($\overline{p}/p$) ratio as a function of {\ppt} for both is nearly one. This explains why the production of anti-particle and particle is almost equal. This leads us to conclude that   
all simulation models (DPMJET-III, Pythia~8, EPOS~1.99, and EPOS-LHC) well describe the experimental data. This statement is also confirmed by the model-to-data ratio shown in the lower panel of fig.~\ref{fig4}.

\begin{figure}[!ht]
\centering
\includegraphics[width=0.49\textwidth,height=0.35\textheight]{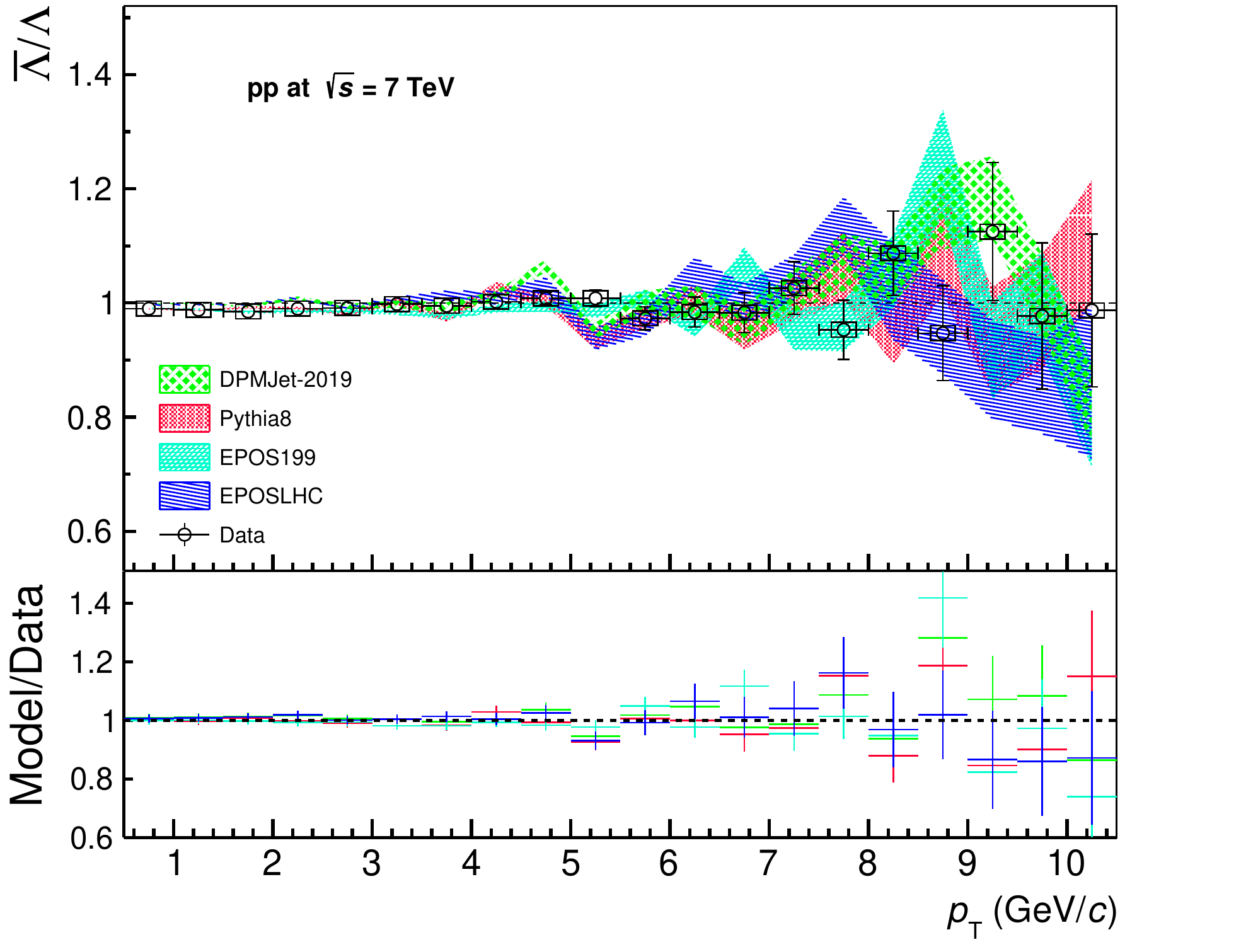}
\includegraphics[width=0.49\textwidth,height=0.35\textheight]{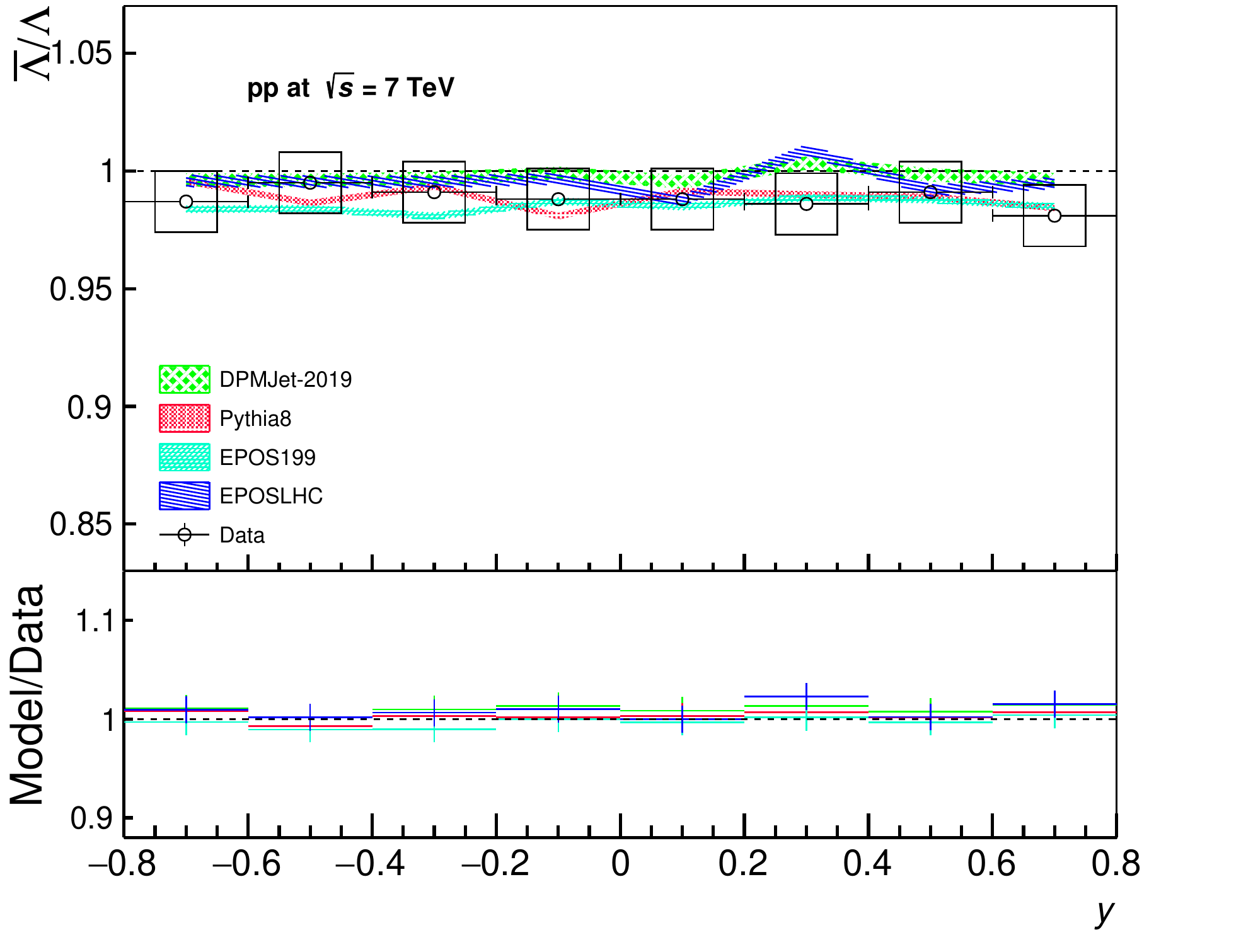}

\caption{{\ppt} dependent {\alam /\lam} ratio (left) and $y$ (right) in $pp$ collisions at {\sqrts} = 7 TeV compared with DPMJET-III, Pythia~8, EPOS~1.99, and EPOS-LHC models. The model-to-data ratio is shown in the lower panel.}
\label{fig5}
\end{figure}

Figure~\ref{fig5} (left) depicts the {\alam /\lam} ratio vs {\ppt} in $pp$ collisions at {\sqrts} = 7 TeV from the ALICE experiment. The experimental data is then compared to simulations calculated from DPMJET-III, Pythia~8, EPOS~1.99, and EPOS-LHC models.  Experimental data in fig.~\ref{fig5} suggests that for low {\ppt} bins ($<$ 5.5 Gev/$c$) the {\alam /\lam} ratio is almost one and at higher {\ppt} bins ($>$ 5.5 Gev/$c$) it is  on average close to unity within uncertainties. A good agreement is observed between model simulations and experimental data including  statistical fluctuations at high {\ppt} bins ({\ppt} $> 7$ GeV/$c$) for both. The model-to-data ratio presented on the lower panel of fig.~\ref{fig5} also shows good agreement less than 5\% difference between experimental data and models for low {\ppt} bins ($<$ 7 GeV/$c$). However, at higher {\ppt} bins ($> 7$ GeV/$c$) the trend deviates from unity upto 40\% for EPOS~1.99 and less than 20\% for all other models. These differences are a direct result of statistical fluctuations in the {\alam /\lam} ratio. {\alam /\lam} ratio as a function of $y$ is shown on fig.~\ref{fig5} (right) which illustrates that model simulations match the experimental data well for all $y$ bins. Additionally, fig.~\ref{fig5} shows that both simulations and experimental data are independent of the rapidity within uncertainties.   

\begin{figure}[!ht]
\centering
\includegraphics[width=0.49\textwidth,height=0.35\textheight]{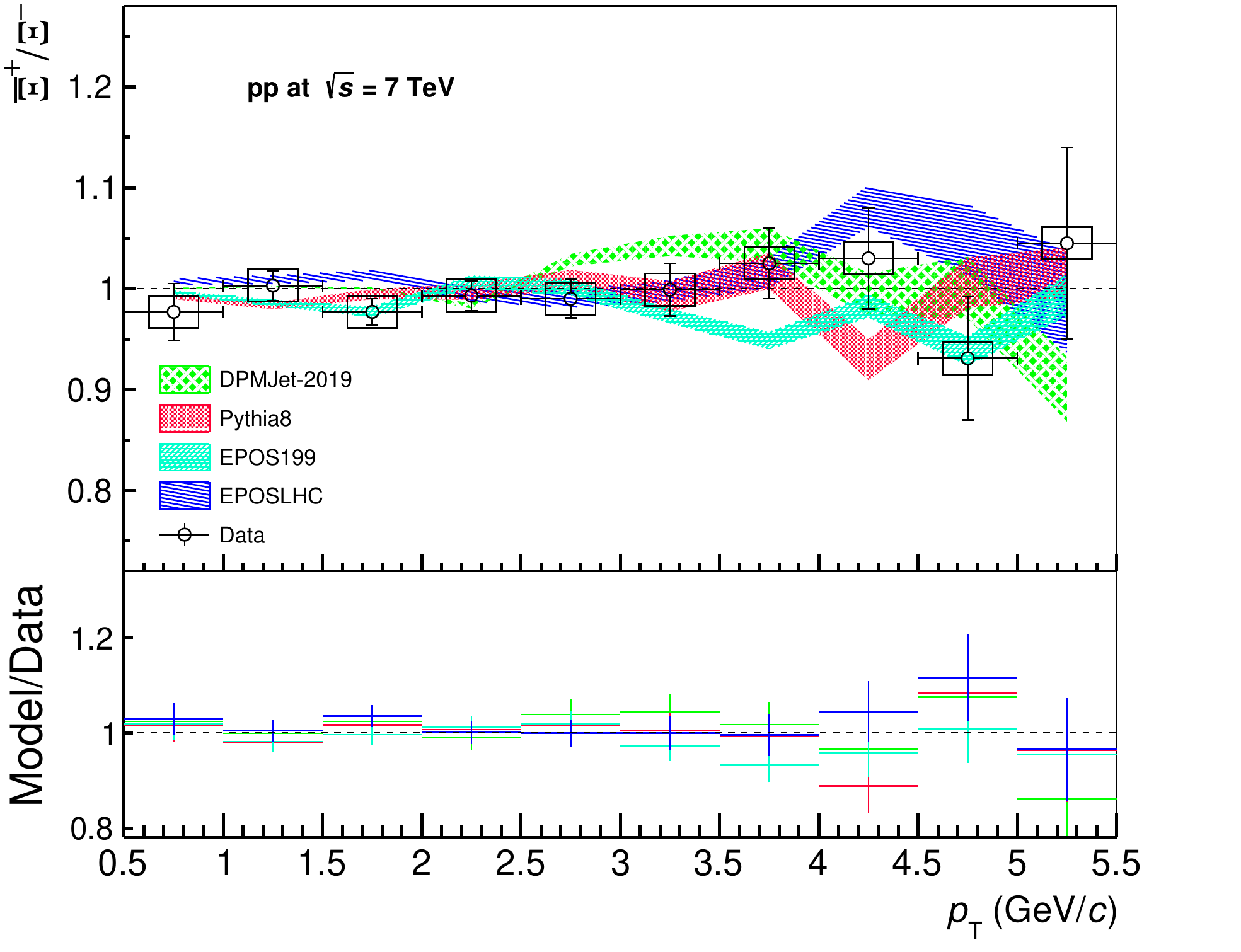}
\includegraphics[width=0.49\textwidth,height=0.35\textheight]{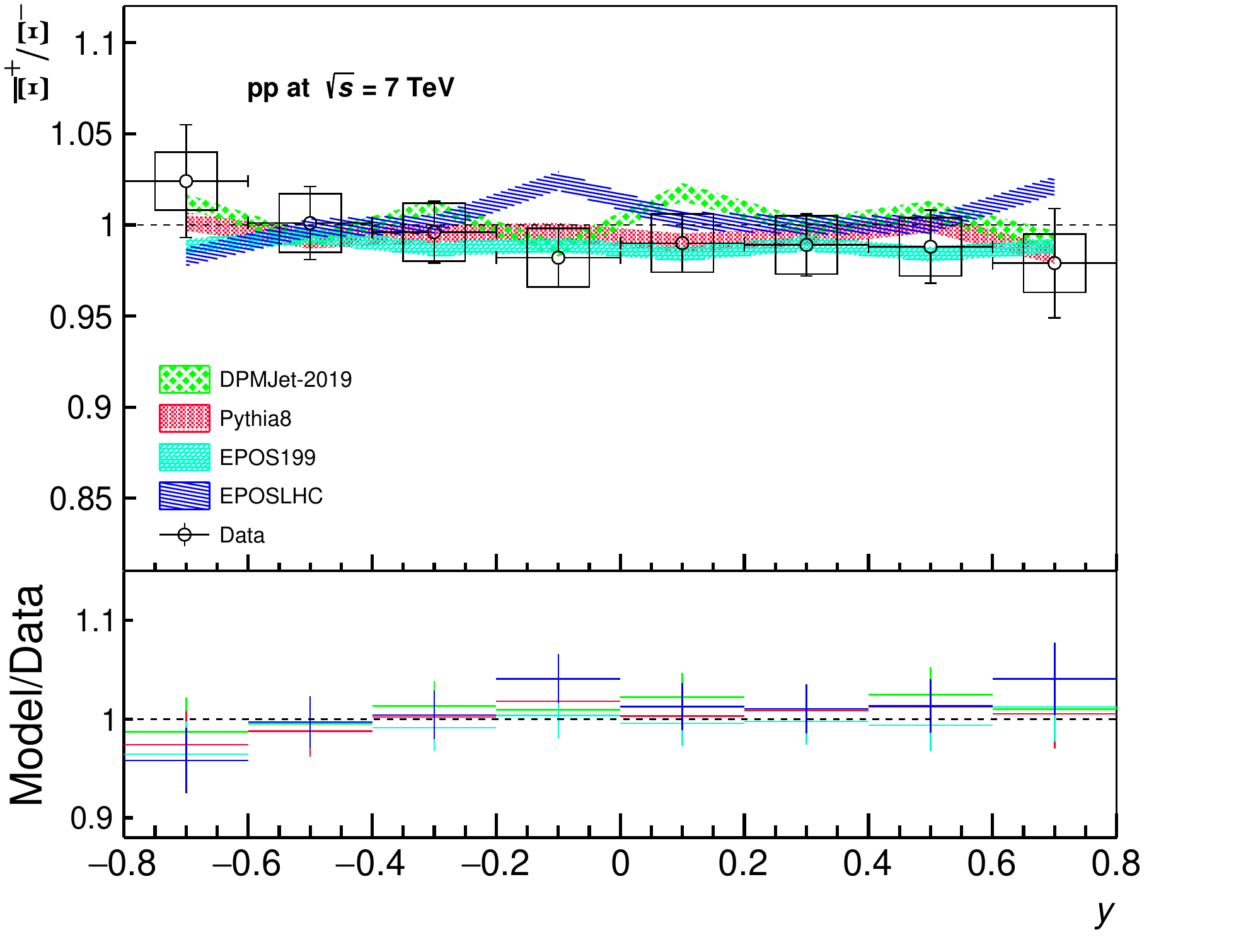}

\caption{{\axi / \xim} ratio vs {\ppt} (left) and $y$ (right) in $pp$ collisions at {\sqrts} = 7 TeV in comparison with DPMJET-III, Pythia~8, EPOS~1.99, and EPOS-LHC models. The model-to-data ratio is shown in the lower panel. }
\label{fig6}
\end{figure}

Figure~\ref{fig6} (left) shows the {\axi / \xim} ratio vs {\ppt} in $pp$ collisions at {\sqrts} = 7 TeV. It is clear from the figure that at lower {\ppt} bins $<$ 4 GeV/$c$ there is good agreement between data and simulations for all models within uncertainties. The model-to-data ratio presented on the lower panel of figure~\ref{fig6} shows the same trend; a good agreement between experimental data and models for low {\ppt} bins $<$ 4 GeV/$c$, and a deviation from unity due to statistical fluctuations for high {\ppt} bins $>$ 4 GeV/$c$ of up to 15\%. The {\axi / \xim} ratio as a function of rapidity is shown in fig.~\ref{fig6} (right). The experimental data and the computed ratio from the different models is consistent with unity within uncertainties for the entire rapidity range. Here again, it is observed that all models give a very good description of experimental data for all $y$ bins.

%As was said in previous sections, overall multiplicity coming from event with presence of different baryon number transport process is proportional to number of broken strings. Due to this, in high multiplicity sample we have higher probability to find events with baryon number transported by string junction itself. Similar explanation can be done for transverse momentum. So in general, antibaryon to baryon ratio can decrease as a function of transverse momentum and/or multiplicity.

\subsection{Discussion}

In this section, we detail the differences between model simulations and the experimental data. We also discuss the changes and updates undertaken in this study to the various model simulations.

There are certain possible ways baryons may be produced in $pp$, $pN$ and $NN$ collisions, that include the pair production of baryon and anti-baryon from a vacuum, which was included in the original Dual Parton Model (DPM). DMP was later updated to include the multi-string fragmentation to enhance the baryon and anti baryon pair production. A good description of the distribution of hadrons in quark and gluon jets can also be given by the string fragmentation process. However, this does not reproduce the baryon stopping exactly and in order to explain the baryon stopping Capella and Kopeliovich (C-K)~\cite{11, 12, 13} suggested the addition of new diquark breaking diagrams, originally introduced by Kharzeev~\cite{14}. The baryon itself can be considered either as a combination of three quarks joined together by a string junction or a combination of diquark and a quark, where the string junction is considered a part of the diquark. This addition of new diquark breaking diagrams required modification to the DPMJET-III model. It should be noted here that for every new diquark produced there needs to be a new parameter introduced into the model. 

%%%%%%%%%%%%%%%%%%%%%%%%%%
\begin{figure}[!ht]
\centering
\includegraphics[width=0.45\textwidth,height=0.3\textheight]{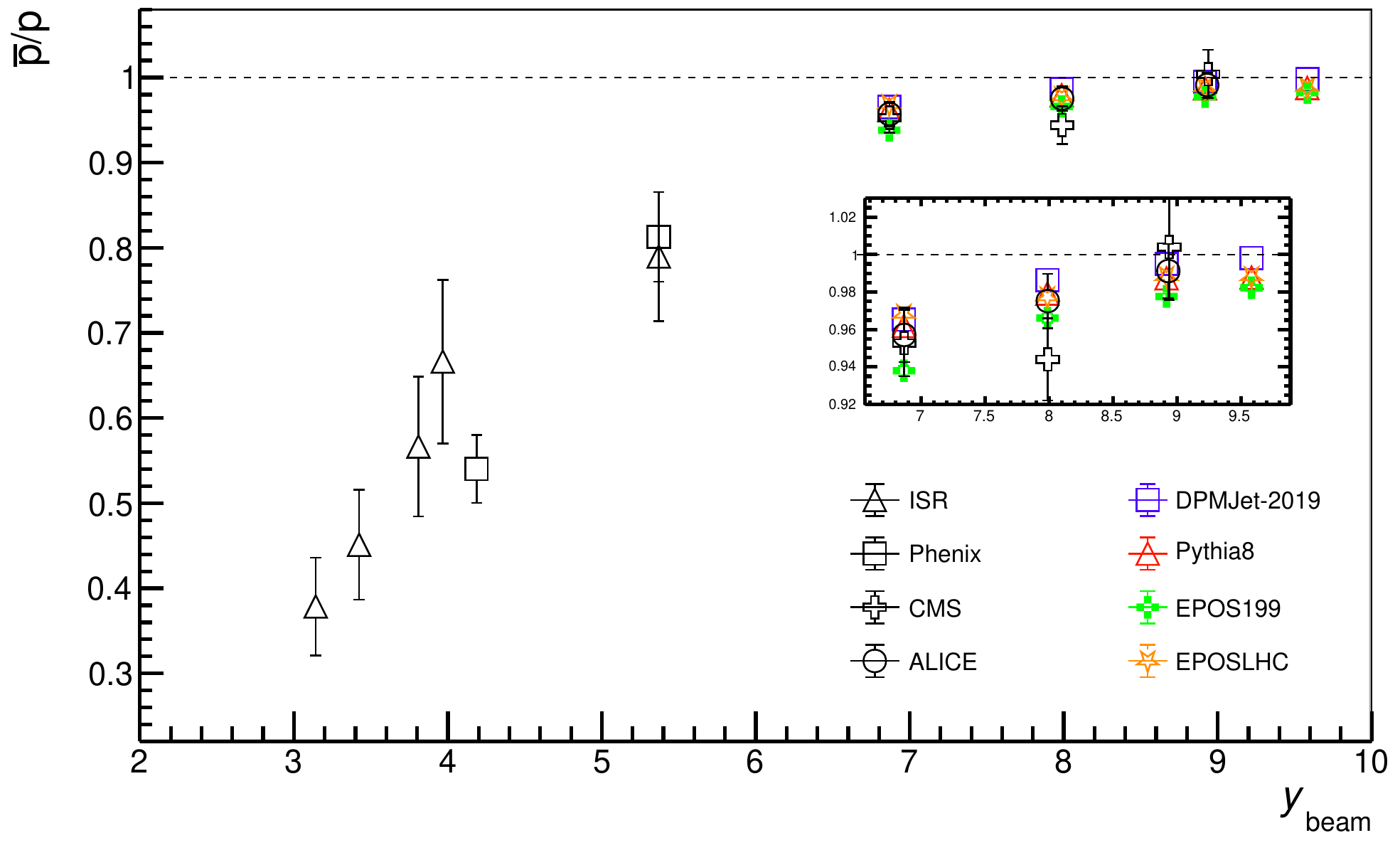}
\includegraphics[width=0.45\textwidth,height=0.3\textheight]{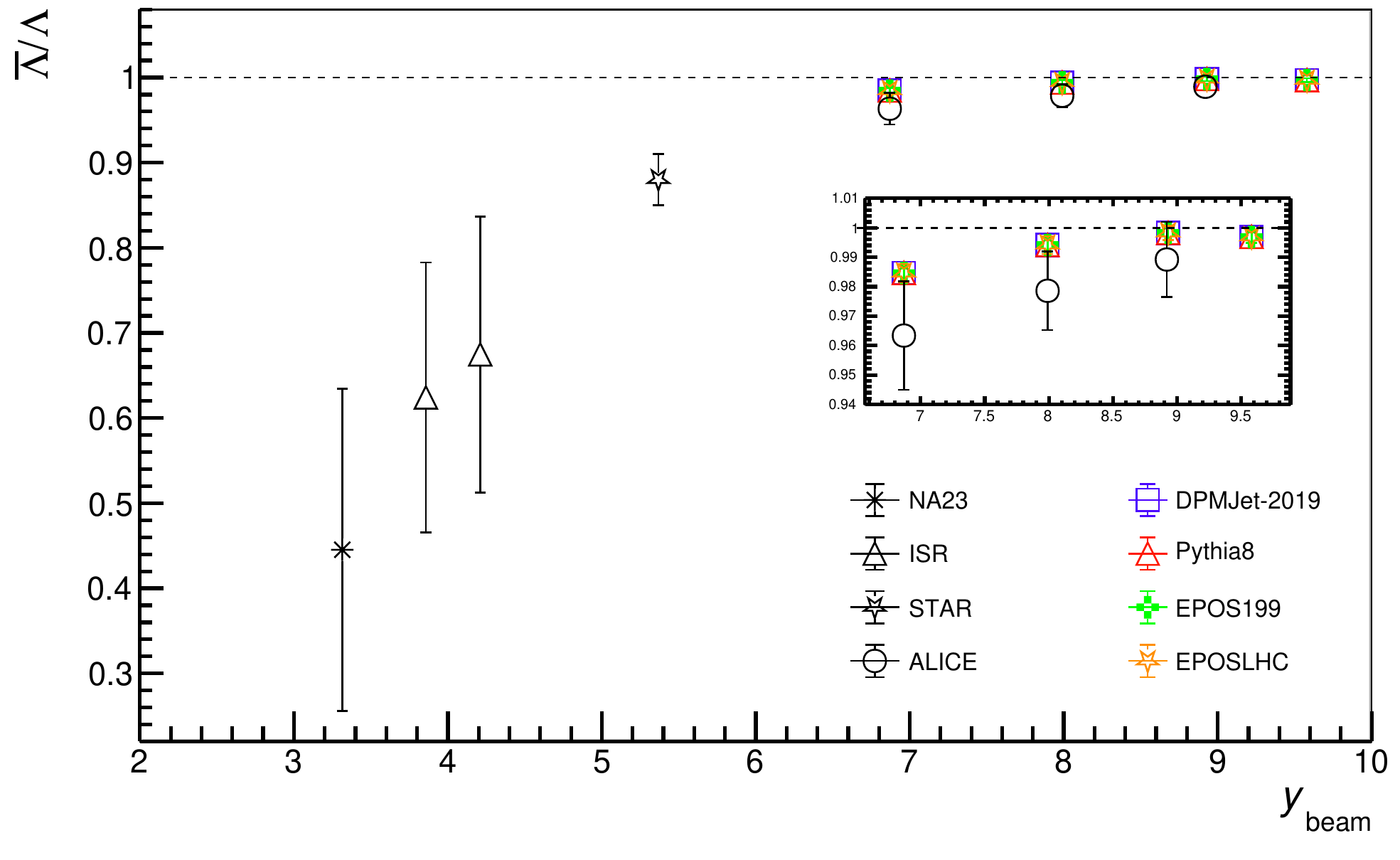}
\includegraphics[width=0.45\textwidth,height=0.3\textheight]{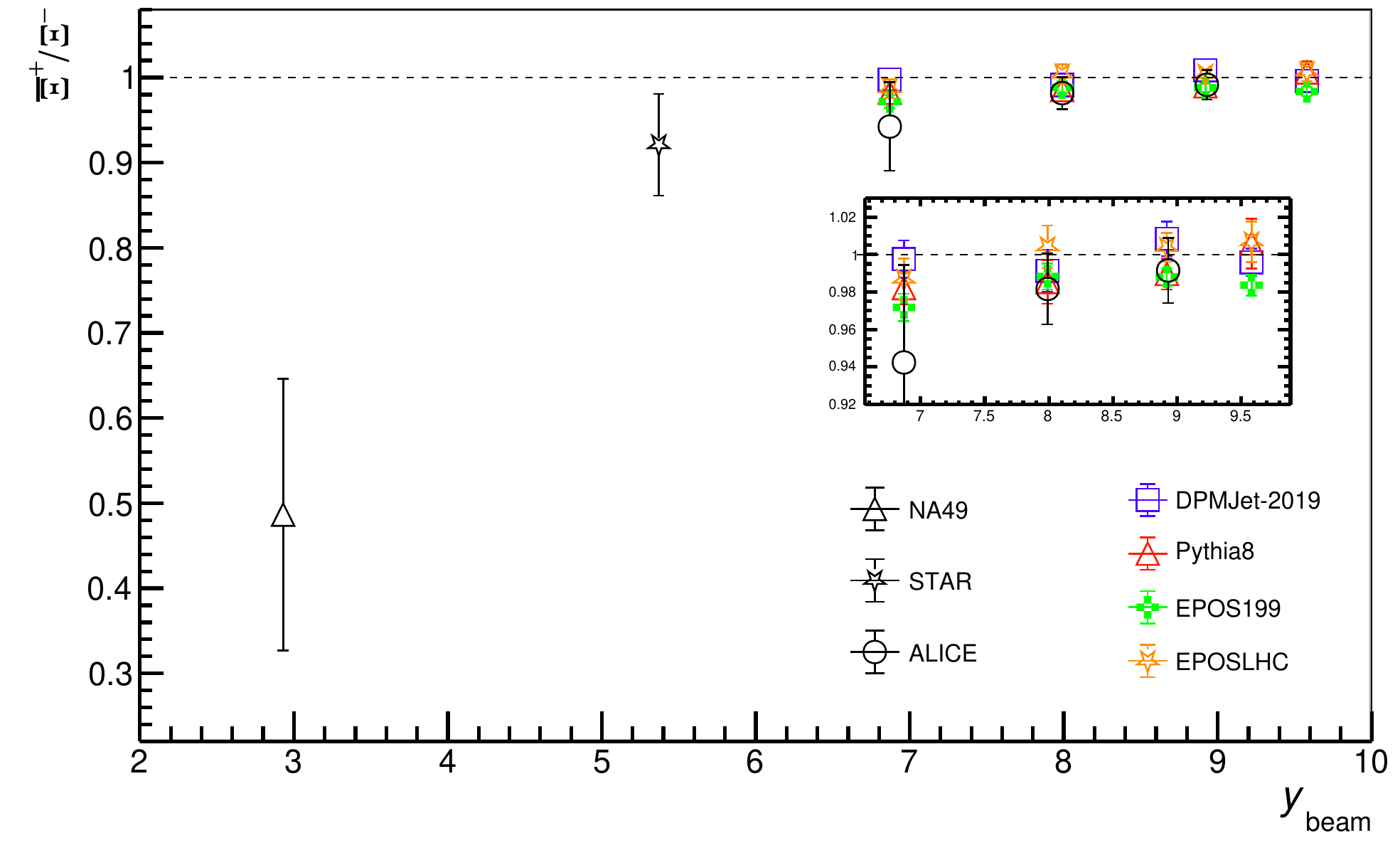}
\caption{$\overline{B}/B$ as a function of $y_{beam}$ in $pp$ {\sqrts} = 0.9, 2.76, 7 and 13.6 TeV in comparison with DPMJET-III, Pythia~8, EPOS~1.99, and EPOS-LHC simulations. Sub-figures are added for better visualization at higher energies. The data at lower rapidity is taken from~\cite{47, 48, 49, 50, 51, 52, 53, 54, 55, 56, 57, 58, 59}.  }
\label{fig8}
\end{figure}
%%%%%%%%%%%%%%%%%%%%%%%%%%%

 The mechanism by which  baryons are produced is more complex than the mechanism by which mesons are produced. As mentioned in section~\ref{sec1} the resulting distribution of the baryon-antibaryon pairs can be understood in terms of the string junction.  The simplest possibility is that the string breaks by the production of a diquark-antidiquark pair, which become the  constituents in the final baryons and the antibaryons. $B\overline{B}$ correlations observed in fig.~\ref{fig8} show that for large rapidity values ($>$ 6)  $B\overline{B}$ are produced in the same ratios, however for low rapidity values ($<$ 6) fewer $\overline{B}$s are produced with respect to $B$s. These results at high rapidity values are in good agreement with the assumption that the $B$ and the $\overline{B}$ are produced together as immediate neighbours in a string breakup.

However, this process gets more complex when we consider the production of a meson along with the production of baryon-antibaryon pairs. This is known as the popcorn fragmentation and is based on the diquark breaking diagrams,  discussed in detail by Ranft~\cite{22}.  Currently, popcorn fragmentation is implemented in the Lund chain fragmentation model in JETSET~\cite{31} and is also used in the DPMJET-III model in this study.

In $NN$ collisions, two baryon stopping mechanisms can take place, the Glauber sea quark mechanism of baryon stopping (GSQBS) and the unitary sea quark mechanism of baryon stopping (USQBS). In GSQBS process fewer $\overline{B}$s are left in the final state than $B$s. A simple explanation of GSQBS is that a valence diquark originating from the target beam and a sea quark originating from the nucleon projectile, join to produce the baryon in latter fragmentation steps. The sea quark in this interaction is called the Glauber sea quark and therefore the mechanism is called the GSQBS. The probability of GSQBS diquark splitting increases if the nucleon interacts more than once. In addition to GSQBS, the baryon stopping mechanism is also a side effect of chain fusion, for example, the fusion of either $qq$-$q$ and $\overline{q}$-$q$ chains to $q$-$qq$ chain~\cite{46}.

USQBS is a special case of the GSQBS that takes place only at high energies and in which a single quark arising from the initial beam remains at the end of the chain. Both GSQBS and USQBS are implemented in  DPMJET-III, which leads to baryon stopping in $pp$ collisions, both at the cosmic ray energies and the collider energies. As mentioned above, for every new diquark produced there needs to be a new parameter introduced in the GSQBS and the USQBS. To determine the optimum values of these new parameters the net-baryon distributions ($\Delta B = B-\overline{B}$) from experimental data are compared to the DPMJET-III model. This study uses the new parameter value = 0.6 for both GSQBS and USQBS. 

Additionally, due to limited experimental information on baryon production, in particular concerning multi-strange baryons, some simplifying approximations are also implemented in the DPMJET-III model. Pythia 8, on the other hand, is a pure multi-parton interaction model. It does not consider junction-stopping mechanisms but rather includes an enhanced baryon transfer from initial to final states. It must be noted here that in Pythia 8, the chain fragmentation pairs of diquark and anti-diquark can be exchanged at any position except close to the end of the chain~\cite{46}. Both EPOS models (EPOS~1.99 and EPOS-LHC) are based on Regge theory, which was introduced in section~\ref{sec1}. EPOS~1.99 is a pre-LHC model designed to simulate energies in the GeV range while LHC energies are in TeV ranges. Hence, EPOS-LHC model provides a comparatively better agreement with experimental data. 

Figure~\ref{fig8} summarises the computed ratios for $\overline{p}/p$, {\alam /\lam} and {\axi / \xim} from various models in comparison with the data from multiple  experiments~\cite{47, 48, 49, 50, 51, 52, 53, 54, 55, 56, 57, 58, 59}, including PHENIX and STAR at BNL and NA23, NA49, ISR, CMS, and ALICE at CERN, as a function of $y_{beam}$. At lower $y_{beam}$, which corresponds to soft scattering, there is an exponential decrease in the probability of production of antibaryons. High $y_{beam} $ implies hard scattering which means there is a large momentum transfer and hence equal numbers of baryons and anti baryons are produced and the $\overline{B}/B$ ratio approaches unity. In order to better visualize the data points from various model simulation a sub figure is included in fig.~\ref{fig8}. We observe that the computed ratios from various models at higher $y_{beam}$ converge to unity and are in good agreement with the experimental data.

 Baryons are identified in experimental measurements by using a higher {\ppt} cut-off value than the mean {\ppt} of the produced baryons~\cite{20}. When comparing the model simulations to data the same {\ppt} were used. Regge models are used to derive the $y$ and $y_{beam}$ dependencies of the $\overline{B}/B$ ratios. It is important to note that in the phenomenological approach, the Pomeron exchange implemented in both EPOS~1.99 and EPOS-LHC models is responsible for the pair production of baryons at very high energies, i.e. at energies greater than current collider energies. 
 
 The string junction transport also expresses the asymmetry ($A$), defined in section~\ref{sec1} as, ($A \equiv \frac{N_{p} - N_{\bar{p}}}{N_{p} + N_{\bar{p}}}$), between the production of baryons versus anti-baryons with an exchange of negative C-parity corresponding to the Reggon ($\omega$) exchange. This asymmetry in $pp$ collisions at {\sqrts} = 0.9, 2.76, 7 and 13.6 TeV, is computed with model simulations and results are provided in table~\ref{tab2}. In conclusion, any significant contribution due to $\omega$ exchange to the $\overline{B}/B$ ratios at mid-rapidity is suppressed. However, this contribution increases with the increase in beam rapidity ($y_{beam}$) resulting in the $\overline{B}/B$ ratio close to one. The asymmetry for protons from table~\ref{tab2}, clearly shows a decreasing trend with increasing energies for all models which means the number of protons and antiprotons are produced in equal numbers at high energies, a trend also confirmed by the predictions of models at {\sqrts} = 13.6 TeV.
%%%%%%%%%%%%%%%%%%%%%%%%%%%%%%%%%%%%%%%%%%%
 %%%%%%%%%%%%%%%%%%%%%%%%%%%%%%%%%%%%%%%%%%%%%%%%%%%%%%%%%%%%%%%%%%%%%
%   This is a LaTeX file.
%%%%%%%%%%%%%%%%%%%%%%%%%%%%%%%%%%%%%%%%%%%%%%%%%%%%%%%%%%%%%%%%%%%%%

%\documentclass[10pt]{article}
%\usepackage{multicol}
%\usepackage{graphicx}
%\usepackage{amsmath}
%\usepackage[a4paper]{geometry}
%\usepackage{rotating}

%\renewcommand{\baselinestretch}{1.05}
%\renewcommand{\thefootnote}{\fnsymbol{footnote}}
%\setlength{\parindent}{.5cm} \setlength{\columnsep}{.5cm}
%\setlength{\oddsidemargin}{-.5cm} \setlength{\topmargin}{-1.5cm}
%\setlength{\textwidth}{17.5cm} \setlength{\textheight}{23.5cm}

%\begin{document}

\begin{table*}[!htb]
{
\begin{center}
\caption{The asymmetry ($A$) of proton computed from model simulations at LHC energies.}

\begin{tabular}{ccccc}\\ \hline \hline
Proton&0.9~TeV&  2.76~TeV & 7~TeV & 13.6~TeV \\
   \hline
     DPMJET-III& 0.0176&	0.0090&	0.0070&	0.0020  \\
     EPOS-LHC& 0.0162	&0.0114&0.0060&	0.0060\\
    EPOS 1.99   &0.0320&	0.0172&	0.0113&	0.0090\\
    Pythia 8 & 0.0195& 0.0104&	0.0098&	0.0061 \\
    \hline\hline
   %&& && \\
    %&&Lambda&&\\
    %&& && \\  

     %
     %DPMJet-2019& 0.008186	&0.00293&	0.0008748&	0.00007117\\
     %EPOS-LHC& 0.0107&	0.005722&	0.002034&	0.00143\\
    
    %EPOS-199   &0.024&	0.0135&	0.007378&	0.005185\\
    %Pythia8 & 0.0139&	0.007775&	0.005579&	0.004826 \\
    %\hline\hline
    % && && \\
    %&& Xi  &&\\
    
    %&& && \\
     
     %DPMJet-2019& 0.001234&	0.004463&	0.004148&	0.00205  \\
    % EPOS-LHC& 0.00673&	0.002024&	0.00186&	0.00335\\
    %EPOS-199   &0.01438	&0.005878&	0.006033&	0.0082\\
    %Pythia8 & 0.00914&	0.006166&	0.0052&	0.00342 \\
     %7   & $|y| < 0.5$   & $|y| < 0.8$    &  $|y| < 0.8$  \\
     %& $0.45 < p_T < 1.05$ GeV/$c$ & $0.5 < p_T < 10.5$ GeV/$c$ &5 $0.5 < p_T < 5.5$ GeV/$c$ \\
     %153.6   & $|y| < 0.5$   & $|y| < 0.8$    &  $|y| < 0.8$  \\
     %55& $0.45 < p_T < 1.05$ GeV/$c$ & $0.5 < p_T < 10.5$ G5eV/$c$ & $0.5 < p_T < 5.5$ GeV/$c$ \\
    
\hline
\end{tabular}
\label{tab2}
\end{center}}
\end{table*}

%\end{document}

\begin{figure}[!ht]
\centering
\includegraphics[width=0.75\textwidth,height=0.35\textheight]{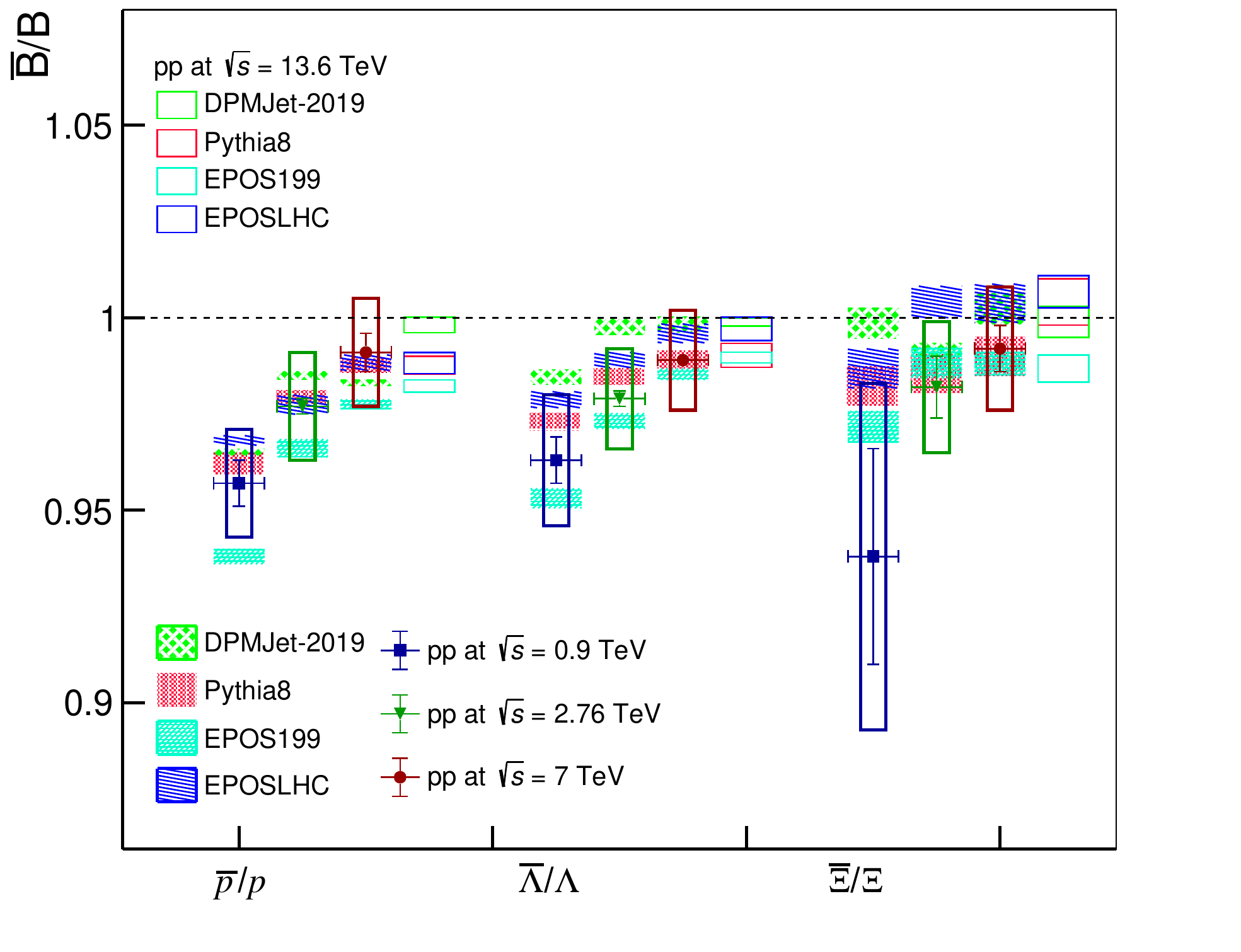}
\caption{ $\overline{B}/B$ ratio at {\sqrts} = 0.9, 2.76 and 7~TeV in comparison with DPMJET-III, Pythia~8, EPOS~1.99, and EPOS-LHC models. Open squares of different colors are the prediction of various models at {\sqrts} = 13.6 TeV. }
\label{fig7}
\end{figure}

Figure~\ref{fig7} shows the measured $\overline{p}/p$, {\alam /\lam} and {\axi / \xim} from the experimental data from ALICE together with the same ratios extracted from DPMJET-III, Pythia~8, EPOS 1.99, and EPOS-LHC models. DPMJET-III models the baryon number stopping mechanism via string-junction transport while Pythia~8 employs a pure multi-parton interaction model. On the other hand, both EPOS 1.99 and EPOS-LHC are based on the Regge theory in which parton ladders split into color strings which then fragment into hadrons. The models used for current study reproduce the data reasonably well, although EPOS 1.99 shoes a steeper rise in the ratio as a function of energy for $\overline{p}/p$ and {\alam /\lam} than the experimental data. Due to the large uncertainties in the experimental data, it is hard to observe an increase of the ratio with strangeness content for the given energy. On the other hand, DPMJET-III and EPOS-LHC models exhibits an increase of the ratio with strangeness content for the given energy and this effect is more prominent in {\alam /\lam} and {\axi / \xim} ratios. For all species, the ratio increases with the increase in energy. 

%%%%%%%%%%%%%%%%%%%%%%%%%%%%%%%%%%%%%%%%%%%
%%%%%%%%%%%%%%%%%%%%%%%%%%%%%%%%%%%%%%%%%%%
             
\section{Summary}\label{sec4}

In this study we computed the $\overline{p}/p$, {\alam /\lam}, {\axi / \xim} ratios from DPMJET-III, Pythia~8, EPOS 1.99, and EPOS-LHC model simulations in $pp$ collisions at {\sqrts} = 0.9, 2.76, 7 and 13.6 TeV. The ratios at 13.6 TeV were computed to check model predictions with experimental data when it is available since LHC is currently taking high luminosity data at 13.6 TeV in Run-3. 

We report that these ratios ($\overline{p}/p$, {\alam /\lam} and {\axi / \xim}) are independent of both the transverse momentum ({\ppt}) and rapidity ($y$). Additionally, when compared to previously published experimental results from ALICE it is observed that all model simulations generally agree with the data. In multiple string fragmentation models, the information of baryon number is carried by the vortex lines which play an important role in the string structure and hence our understanding of baryons and particularly net baryon production ($\Delta B = B - \overline{B}$). The $\overline{B}/B$ ratios approach unity as mass of the constituent quarks increases. Additionally, it is also observed that, DPMJET-III and EPOS-LHC models exhibits an increase of the ratio with strangeness content for the given energy and this effect is more prominent in {\alam /\lam} and {\axi / \xim} ratios. This effect is not prominent in the experimental data due to large uncertainties. For all species, the ratio increases with the increase in energy similar to the experimental data. 

At lower energies, we observe an excess of baryons over anti-baryons in all model simulations, which is directly related to the baryon number transfer from the incoming beam. However, this is not the case at higher energies, where baryon and anti-baryons are produced in equal numbers because of the baryon-and-anti-baryon pair production. This trend is somewhat alleviated by the introduction of the new diquark breaking diagrams in DPMJET-III model, which is why DPMJET-III shows the best correlation with experimental as well as the prediction at 13.6 TeV. The model simulations also suggest a saturation of pair production at higher energies leading to the ($\overline{B}/B$) ratio approaching unity. Model simulations also predict similar spectra, for both {\ppt} and $y$, for baryons and anti-baryons at large rapidity in $pp$ collisions. 

We also report the predictions of asymmetry ($A$) for protons from the model simulations in $pp$ collisions at all {\sqrts} = 0.9, 2.76, 7.0 and 13.6 TeV. Previously the asymmetry was reported only for DPMJET-II, so our results will help put additional constraints on model calculations. The asymmetry shows a decreasing trend at higher energies for all model simulations, which is consistent with previously available experimental measurements.

\vskip0.5cm
\textbf{Acknowledgement:}
 The authors would like to show our deepest gratitude to the following individuals, Mr. Muhammad Salman Ashraf from the  Institute of Space Technology, Islamabad, and Dr. Muhammad Ahmad from the National Centre for Physics, Islamabad, and O\u{g}uz G\"{u}zel Centre for Cosmology, Particle Physics and Phenomenology (CP3), Université Catholique de Louvain, B-1348 Louvain-la-Neuve, Belgium, for sharing their pearls of wisdom and healthy discussions during the course of this research. 

\vskip0.5cm
\textbf{Availability of Data and Material:}
The authors declare that all the supported data of this study are available within the article.

%\endpage
%\begin{thebibliography}{}
%\bibitem{}
\bibliographystyle{unsrt}
%
%\end{thebibliography}

%\end{multicols}

\end{document}